\documentclass[final,5p,times,twocolumn]{elsarticle}

\usepackage{amsmath}
\usepackage{graphicx}
\usepackage{txfonts}
\usepackage{multirow}
\usepackage{verbatim}
\usepackage{array}
\usepackage{threeparttable}
\usepackage{algorithm}
\usepackage{algorithmic}
\usepackage{paperSyml}

\newtheorem{definition}{Definition}

\journal{arXiv}

\begin{document}

\begin{frontmatter}



\title{Triangle-oriented Community Detection considering Node Features and Network Topology}

 \author[1]{Guangliang Gao}
 \author[2]{Weichao Liang}
 \author[1]{Ming Yuan}
 \author[1]{Hanwei Qian}
 \author[1]{Qun Wang}
 \author[3]{Jie Cao\corref{cor}}
 \ead{caojie690929@163.com}
 
 \cortext[cor]{Corresponding author}
 
 \affiliation[1]{organization={Department of Computer Information and Cyber Security, Jiangsu Police Institute},
             city={Nanjing},
             country={China}}

 \affiliation[2]{organization={School of Computer Science and Engineering, Nanjing University of Science and Technology},
             city={Nanjing},
             country={China}}
         
 \affiliation[3]{organization={Jiangsu Provincial Key Laboratory of E-Business, Nanjing University of Finance and Economics},
         	city={Nanjing},
         	country={China}}

\begin{abstract}
The joint use of node features and network topology to detect communities is called community detection in attributed networks. Most of the existing work along this line has been carried out through objective function optimization and has proposed numerous approaches. However, they tend to focus only on lower-order details, i.e., capture node features and network topology from node and edge views, and purely seek a higher degree of optimization to guarantee the quality of the found communities, which exacerbates unbalanced communities and free-rider effect. To further clarify and reveal the intrinsic nature of networks, we conduct triangle-oriented community detection considering node features and network topology. Specifically, we first introduce a triangle-based quality metric to preserve higher-order details of node features and network topology, and then formulate so-called two-level constraints to encode lower-order details of node features and network topology. Finally, we develop a local search framework based on optimizing our objective function consisting of the proposed quality metric and two-level constraints to achieve both non-overlapping and overlapping community detection in attributed networks. Extensive experiments demonstrate the effectiveness and efficiency of our framework and its potential in alleviating unbalanced communities and free-rider effect.
\end{abstract}




\end{frontmatter}



\section{Introduction}\label{sect1}
Networks provide a natural way to express the complex relationships in our daily life, such as scientific collaborations~\cite{alshebli2018preeminence}, friend interactions~\cite{Liu9271901}, information dissemination~\cite{liu2022eriskcom}, and module associations~\cite{meng2021protein}. Typically, the individuals and their relationships in real-world scenarios are represented as the nodes and edges in networks where each node is associated with one or more features characterizing the properties of the individual it corresponds. More formally, we call the networks following the above way of expression as attributed networks~\cite{falih2018community, DBLP:journals/csr/Chunaev20}. The goal of community detection in attributed networks is to identify the communities hidden inside the networks by utilizing the node features and network topology comprehensively, which not only helps us get a deeper understanding of the network structure, but also gives the people new insights into a series of related issues~\cite{DBLP:journals/pr/ZhangWZC17, DBLP:conf/wsdm/DingLL19, DBLP:conf/ijcai/LiuX0ZHPNYY20, su2022comprehensive}.

Over the past several decades, objective function optimization has shown its significance in detecting communities in attributed networks, thus attracting a great deal of attention from numerous researchers~\cite{DBLP:journals/pvldb/ZhouCY09, DBLP:conf/kdd/ZheSX19, jia2017node, DBLP:conf/icdm/YangML13, DBLP:journals/inffus/BuGLC17, DBLP:journals/tcyb/LiLW18}. Roughly speaking, existing studies in this field can be categorized into two families based on the strategies they employ in integrating node features and network topology. The former kind of approaches address the node features and network topology sequentially. For instance, SA-Cluster~\cite{DBLP:journals/pvldb/ZhouCY09}, AGGMMR~\cite{DBLP:conf/kdd/ZheSX19}, and $k$NN-enhance~\cite{jia2017node} first embed the node features into the network topology adopting the way like adding (weighting) nodes or edges, and then deal with the objective optimization problem on the augmented network. In contrast, the latter one handles the node features and network topology simultaneously. For example, CESNA~\cite{DBLP:conf/icdm/YangML13}, CAMAS~\cite{DBLP:journals/inffus/BuGLC17}, and MOEA-SA~\cite{DBLP:journals/tcyb/LiLW18} first define the objective functions based on the node features and network topology respectively, and then find a balance of optimization through methods such as a trade-off parameter, multi-objective optimization, and non-negative matrix factorization.

Although the above methods have been proven successful in many cases, there remain some issues that deserves careful consideration when performing community detection in attributed networks. Firstly, most of the approaches focus on lower-order details only, i.e., capturing the node features and network topology from the node and edge views. Studies~\cite{benson2016higher, DBLP:conf/wsdm/ParanjapeBL17, DBLP:conf/www/RossiAK18} have demonstrated that higher-order details, i.e., motifs, are conducive to uncovering the underlying mechanisms of networks. Though there are some attempts~\cite{DBLP:conf/kdd/YinBLG17, DBLP:conf/icde/HuangLBL21, xia2021chief} which detect communities based on motifs, they are often not applicable to attributed networks. Secondly, it is a consensus~\cite{falih2018community, DBLP:journals/csr/Chunaev20, DBLP:journals/pvldb/ZhouCY09, DBLP:conf/kdd/ZheSX19, jia2017node, DBLP:conf/icdm/YangML13, DBLP:journals/inffus/BuGLC17, DBLP:journals/tcyb/LiLW18} that we wish the nodes belonging to the same community are with dense edges and homogeneous features while the nodes falling into different communities are not. However, it is unwise to focus on nothing but the optimal value of the objective function to achieve the above goal, since it may lead the communities we find suffer from overload, unbalance~\cite{DBLP:journals/tsmc/GaoWZCQ22} or free-rider effect~\cite{DBLP:journals/pvldb/WuJLZ15}. Thirdly, scalability, ignored by many researchers, is an important factor to consider. Optimization based on global models~\cite{DBLP:conf/icdm/YangML13, DBLP:conf/pakdd/GunnemannBFS13, DBLP:conf/ida/CombeLGE15} is computational expensive, especially under the case with large-scale features and complex topology. Precisely for this reason, many of the existing approaches are impossible to be used for many tasks.

In this study, we still follow the principle of objective function optimization and try to improve the quality of the found communities through addressing the above-mentioned issues appropriately. Specifically, we first introduce a quality metric to preserve the higher-order details of the node features and network topology by making a trade-off between them in accordance with the information from the network. Then, we define two constraint items to factor into the lower-order details of the node features and network topology for the purpose of alleviating the unbalance and free-rider effect which may occur in the found communities. Finally, an optimization scheme is designed from a local perspective to make our method have high efficiency. The specific contributions of our work are listed as follows:

\begin{itemize}
	\item{We propose a parameter-free quality metric based on the concept of closed topology and feature triangles, which not only evaluates the quality of higher-order structures, but can also be treated as an optimization objective to achieve triangle-based community detection.}
	
	\item{We formulate the so-called two-level constraints from the node and edge perspectives to enhance the capability of the proposed metric as an optimization objective, which further improves the topological tightness and feature homogeneity of each community found.}
	
	\item{We develop a local search framework based on optimizing our objective function consisting of the proposed metric and two-level constraints, which effectively and efficiently reveals both non-overlapping and overlapping community structures in attributed networks.}
\end{itemize}

The remainder of this study is organized as follows. We review the related work in Section~\ref{sect2}, and Section~\ref{sect3} formulates the problem of community detection in attributed networks and illustrates the quality metric referenced in this study. A detailed description of our methodology is presented in Section~\ref{sect4}, and Section~\ref{sect5} shows our experimental results. Finally, we conclude this study and give guidelines for our future work in Section~\ref{sect6}.

\section{Related Work}\label{sect2}
Since node features and network topology are two different kinds of information, community detection in attributed networks aims to make them complement each other to identify high quality communities. Existing approaches that follow objective optimization can be divided into three groups. In this section, we give a brief review of their recent advances and then discuss how our study differs from theirs.

The first group is node-oriented approaches~\cite{DBLP:journals/pvldb/ZhouCY09, DBLP:conf/sigmod/XuKWCC12, DBLP:journals/tcyb/BuLCWG19, xu2022attributed, DBLP:journals/fgcs/CaoWJD19}, which focus on formulating various distance or similarity functions to find communities through attributed network clustering. For instance, Zhou et al.~\cite{DBLP:journals/pvldb/ZhouCY09} developed a clustering framework SA-Cluster, which uses a unified distance metric to measure both topological and feature similarity~\cite{DBLP:journals/ijcat/WuCA12}, and follows the clustering process of \textit{K}-medoids. Xu et al.~\cite{DBLP:conf/sigmod/XuKWCC12} transformed attributed network clustering into a standard probabilistic inference problem based on a defined Bayesian probabilistic model~\cite{DBLP:journals/eswa/WuPZCZZ15} and proposed a variational algorithm to solve it, which avoids the artificial design of distance functions. Bu et al.~\cite{DBLP:journals/tcyb/BuLCWG19} formalized attributed network clustering as a dynamic cluster formation game, and found a balanced solution by designing the feasible action set, the utility function, and the self-learning strategy for each node. Xu et al.~\cite{xu2022attributed} first built an attributed network embedding framework and adopted a distributed algorithm to obtain the embedding vector of each node, and then automatically determined the number of found communities based on curvature and modularity. Finally, the community detection results are obtained by clustering the embeddings. Cao et al.~\cite{DBLP:journals/fgcs/CaoWJD19} proposed an NMF-based model combining node features and network topology, which employs graph regularization to penalize the dissimilarity of nodes and introduces so-called $K$-near neighbor consistency to recover feature information.

The second group is edge-oriented approaches~\cite{DBLP:journals/tkdd/SmithZLP16, malhotra2021modified, BERAHMAND20221869, DBLP:conf/kdd/ZheSX19, DBLP:journals/ijon/XieSLZL21}. Since community detection considering network topology has been widely studied~\cite{DBLP:conf/ijcnn/LiuWZY19, DBLP:journals/www/LiuWXZYS20}, an intuitive strategy is to enhance the network topology with node features, and then find communities based on the enhanced topology. For instance, Smith et al.~\cite{DBLP:journals/tkdd/SmithZLP16} introduced node features into the description of information flow, and then fine-tuned the Infomap algorithm to find communities with large information flow among nodes and similar node features. Malhotra and Chug~\cite{malhotra2021modified} designed four variants based on the label propagation algorithm that exploit node features and edge strength to improve the quality of found communities and overcome the random community allocation problem. Berahmand et al.~\cite{BERAHMAND20221869} also considered the label propagation algorithm. First, a weighted network combining node features and network topology is generated. Then the influence of each node is calculated using Laplacian centrality, thus enhancing the update path of community labels. Zhe et al.~\cite{DBLP:conf/kdd/ZheSX19} developed a three-stage framework AGGMMR consisting of augmented graph construction and weight initialization, weight learning with modularity maximization, and modularity refinement to find communities in attributed networks. Xie et al.~\cite{DBLP:journals/ijon/XieSLZL21} defined a scoring function to check the properties and influence of communities and developed two community search algorithms by maximizing it, and then designed a graph refining algorithm and pruning rules to ensure search efficiency.

The third group is motif-oriented approaches~\cite{DBLP:conf/kdd/YinBLG17, DBLP:conf/icde/HuangLBL21, xia2021chief, DBLP:conf/kdd/SotiropoulosT21, DBLP:conf/kdd/LiHWL19, DBLP:journals/access/LiHWHL18, DBLP:journals/icae/HuPYHH21}. Motifs lie between the microscopic proximity structure and mesoscopic community structure and help find communities that maintain building blocks in the network. Recent studies~\cite{DBLP:conf/kdd/YinBLG17, DBLP:conf/icde/HuangLBL21, xia2021chief, DBLP:conf/kdd/SotiropoulosT21, DBLP:conf/kdd/LiHWL19} usually only consider network topology and rarely involve node features. For instance, Huang et al.~\cite{DBLP:conf/icde/HuangLBL21} studied the motif-based graph partitioning (MGP) problem. First, a sampling-based MGP (SMGP) framework is designed, which employs an unbiased sampling mechanism to estimate edge weights. Furthermore, an adaptive sampling framework SMGP+ is proposed, which adaptively adjusts the sampling distribution and iteratively partitions the input graph based on up-to-date estimated edge weights. Sotiropoulos and Tsourakakis~\cite{DBLP:conf/kdd/SotiropoulosT21} demonstrated the power edge triangle counts for spectral sparsification and advanced the understanding of triangle-based graph partitioning by empirically analyzing two heuristics for community detection. Li et al.~\cite{DBLP:conf/kdd/LiHWL19} proposed an edge enhancement method for motif-aware community detection in purely topological networks. For attributed networks, they~\cite{DBLP:journals/access/LiHWHL18} first formulated an AHMotif adjacency matrix to encode node features and network topology from a higher-order perspective, and then utilized proximity-based methods to find communities. Hu et al.~\cite{DBLP:journals/icae/HuPYHH21} composed tensors to model higher-order patterns in terms of node features and network topology, and developed a novel algorithm to capture these patterns to find communities.

Although community detection in attributed networks through objective optimization has been extensively studied, our study is significantly different from most of the existing work in the following respects. First, we consider both node features and network topology based on observations of the nature of networks rather than pursuing an elaborate integration strategy. Second, we exploit node features and network topology from both higher- and lower-order perspectives, namely closed topology and feature triangles, and two-level constraints. Third, our proposed local search framework is suitable for different community detection tasks. To sum up, our study guarantees the robustness of the process of community detection and shows better effectiveness and efficiency.

\section{Preliminaries}\label{sect3}
In this section, we first formulate the problem of community detection in attributed networks and give the notations used throughout this study. Then, we introduce the quality metric weighted community clustering ($WCC$).

\subsection{Problem Formulation}
Consider representing an attributed network as an undirected and unweighted graph ${\GG}$ = (${\VV}$, ${\EE}$, ${\FF}$), where ${\VV}$ = \{${\vv}_{1}$, ${\vv}_{2}$, ..., ${\vv}_{\nn}$\} is a set of ${\nn}$ nodes, ${\EE}$ ${\subseteq}$ ${\VV}$ ${\times}$ ${\VV}$ is a set of ${\mm}$ edges that connect two nodes of ${\VV}$, and ${\FF}$ = \{$\textbf{\ff}_{1}$, $\textbf{\ff}_{2}$, ..., $\textbf{\ff}_{\nn}$\} is a set of feature vectors associated with the nodes in $\VV$. For any node ${\vv}_{\ii}$ $\in$ ${\VV}$, the neighborhood of ${\vv}_{\ii}$ is the set ${\NN} ({\vv}_{\ii})$ = \{${\vv}_{\jj}$ ${\in}$ ${\VV}$ $|$ (${\vv}_{\ii}$, ${\vv}_{\jj}$) ${\in}$ ${\EE}$\}, the degree of ${\vv}_{\ii}$ is defined as ${\dd}_{\ii}$ = $|{\NN} ({\vv}_{\ii})|$, and $\textbf{\ff}_{i}$ $\in$ $\mathbb{R}^{1{\times}\pp}$ denotes the feature vector of $\vv_{i}$, where $\pp$ is the dimension of the feature vector. We also use the adjacency matrix $\textbf{\AAA}$ ${\in}$ $\{0, 1\}_{\nn {\times} \nn}$ and the node feature matrix $\textbf{\BBB}$ $=$ $\{\BBB_{\ii\jj}\}$ $\in$ $\mathbb{R}^{\nn{\times}\pp}$ to represent the topology and the attributes of graph ${\GG}$, respectively. Thus, an edge (${\vv}_{\ii}$, ${\vv}_{\jj}$) ${\in}$ ${\EE}$, ${\AAA}_{\ii\jj}$ = ${\AAA}_{\jj\ii}$ = 1; otherwise, ${\AAA}_{\ii\jj}$ = ${\AAA}_{\jj\ii}$ = 0. If the $\jj$th feature is presented in $\textbf{\ff}_{i}$ of node ${\vv}_{\ii}$, then ${\BBB}_{\ii\jj}$ $\in$ $(0, 1]$; otherwise, ${\BBB}_{\ii\jj}$ = 0. Note that our discussion is not limited to binary features, but also continuous-valued features.

Community detection in graph $\GG$ aims to find a partition ${\CC}$ = \{${\CC}_{1}, {\CC}_{2}, ..., {\CC}_{\KK}$\} of its nodes such that ${\VV}$ = ${\bigcup}_{\kk = 1}^{\KK}{\CC}_{\kk}$ and a certain balance between the following two objectives is achieved:

Tightness, i.e., a group of nodes have a high density of edges within them, and a lower density of edges between groups.

Homogeneity, i.e., a group of nodes have similar feature values within them, and may have diverse feature values between groups.

\subsection{Quality Metric: $WCC$}
In general, the probability of closed topological triangles among nodes in the same community is larger than the expected among nodes in different communities. $WCC$~\cite{Prat2014WWW, DBLP:conf/icdm/LyuBZZ16} is inspired by this to measure the quality of a partition of nodes. Given the topology ($\VV$, $\EE$) of graph $\GG$, the degree of belonging of a node ${\vv}_{\ii}$ to a community ${\CC}_{k}$, namely $	WCC({\vv}_{\ii},{\CC}_{k})$, is defined as follows:
\begin{equation}
	\label{equ31}
	\left\{
	\begin{split}
		&\frac{t({\vv}_{\ii}, {\CC}_{k})}{t({\vv}_{\ii}, {\VV})}{\cdot}
		\frac{vt({\vv}_{\ii}, {\VV})}{|{\CC}_{k} - \{{\vv}_{\ii}\}| + vt({\vv}_{\ii}, {\VV} - {\CC}_{k})},
		~t({\vv}_{\ii}, \VV) \neq 0; \\
		&0,~~~~~~~~~~~~~~~~~~~~~~~~~~~~~~~~~~~~~~~~~~~~~~~~~~~~~~~~t({\vv}_{\ii}, \VV) = 0; \\
	\end{split}
	\right.
\end{equation}
where $t({\vv}_{\ii}, {\CC}_{k})$ and $t({\vv}_{\ii}, {\VV})$ mean the number of closed topological triangles composed of node ${\vv}_{\ii}$ and the nodes in ${\CC}_{k}$ and ${\VV}$, respectively. $vt({\vv}_{\ii}, {\VV})$ and $vt({\vv}_{\ii}, {\VV} - {\CC}_{k})$ mean the number of nodes in ${\VV}$ and ${\VV} - {\CC}_{k}$ that form at least one closed topological triangle with node ${\vv}_{\ii}$, respectively. $|{\CC}_{k} - \{{\vv}_{\ii}\}|$ means the size of ${\CC}_{k}$ except node ${\vv}_{\ii}$.

Then, the $WCC$ score of a community ${\CC}_{k}$ is defined as follows:
\begin{equation}
	\label{equ32}
	WCC({\CC}_{k}) = \frac{1}{|{\CC}_{k}|}\sum_{\forall {\vv}_{\ii} \in {\CC}_{k}}WCC({\vv}_{\ii},{\CC}_{k}).
\end{equation}

Finally, for a partition ${\CC}$ = \{${\CC}_{1}, {\CC}_{2}, ..., {\CC}_{\KK}$\}, the $WCC$ score is defined as follows:
\begin{equation}
	\label{equ33}
	WCC({\CC}) = \frac{1}{|\VV|}\sum_{k=1}^{\KK}(|{\CC}_{k}|~{\cdot}~WCC({\CC}_{k})).
\end{equation}

It is obvious that this metric is suitable for both non-overlapping and overlapping community detection, and a higher score means better community structure found.

\section{Methodology}\label{sect4}
In this section, we first extend the $WCC$ metric by introducing the concept of closed feature triangles. Then, we describe the proposed tightness and homogeneity constraints, which can improve the capability of the extended $WCC$ as an optimization objective. Finally, we design a local search framework to achieve both non-overlapping and overlapping community detection by maximizing the objective function consisting of the extended $WCC$ and the above constraints.

\begin{table*}[t!]
	\centering
	\begin{threeparttable}
		\caption{Statistics of Edges in Closed Feature Triangles in \textit{Facebook} and \textit{Sinanet} Ground-truths}
		\begin{tabular}{| c | c | c | c | c | c |}
			\hline
			\multirow{2}{*}{Network}	&\multirow{2}{2.0cm}{Total number \\ of triangles\tnote{1}} &\multicolumn{4}{c|}{Number of triangles with different number of edges}\\
			\cline{3-6}
			&	&no edge	&one edge	&two edges	&three edges\\
			\hline
			\textit{Facebook}	&30, 738, 546	&13, 748, 452	&10, 110, 887	&3, 803, 458	&3, 075, 749\\
			\hline
			\textit{Sinanet}	&138, 407, 278	&137, 037, 199	&1, 307, 192	&60, 282	&2, 605\\
			\hline
		\end{tabular}
		
		\begin{tablenotes}
			\footnotesize
			\item[1] Those closed feature triangles are formed by nodes belonging to the same community.
		\end{tablenotes}
		
		\label{tab41}
	\end{threeparttable}
\end{table*}

\begin{table}[t!]
	\centering
	\begin{threeparttable}
		\caption{Number of Closed Topology and Feature Triangles in \textit{Facebook} and \textit{Sinanet} Ground-truths}
		\begin{tabular}{| c | c | c |c |}
			\hline
			\multirow{2}{*}{Network}	&\multirow{2}{1.2cm}{Type of \\ triangle}	&\multicolumn{2}{c|}{Number of triangles}\\
			\cline{3-4}
			&	&ground-truths\tnote{1}	&all communities\tnote{2}\\
			\hline
			\multirow{2}{*}{\textit{Facebook}}	
			&Topology	&1, 209, 670	&1, 125, 137\\
			\cline{2-4}
			&Feature	&1, 007, 762, 916	&30, 738, 546\\
			\hline
			\multirow{2}{*}{\textit{Sinanet}}	
			&Topology	&35, 882	&5, 915\\
			\cline{2-4}
			&Feature	&208, 039, 606	&138, 407, 278\\
			\hline
		\end{tabular}
		
		\begin{tablenotes}
			\footnotesize
			\item[1] Those nodes that form the triangle are in the ground-truths.
			\item[2] Those nodes that form the triangle belong to the same community in the ground-truths.
		\end{tablenotes}
		
		\label{tab42}
	\end{threeparttable}
\end{table}

\subsection{Extension of $WCC$}
Triangles, as fundamental paths and motifs that recur in real-world networks, could be used to define and identify communities and more general classes of nodes. $WCC$ only focuses on the properties of communities from the perspective of closed topological triangles. In fact, there is a consensus from extensive research~\cite{falih2018community, DBLP:journals/csr/Chunaev20} that node feature information can be treated as a supplement to explain the formation mechanism of communities as it becomes available. Hence, we define the closed feature triangle based on the intuition of $WCC$ as follows:
\begin{definition}\label{def31}
A triangle is called the \textbf{closed feature triangle} if there exists at least one feature dimension such that for any three nodes $\vv_{x}, \vv_{y}, \vv_{z} \in \VV$, the following condition is satisfied:
	
In term of binary features, we directly compare the feature values, i.e., $\exists l \in \{1, 2, ..., p\},~B_{xl} = B_{yl} = B_{zl} = 1$.

In term of continuous-valued features, we intuitively consider the feature dimension with the largest value, i.e., $\exists l \in \{1, 2, ..., p\},~\max B_{xl}\not= 0 \wedge \max B_{yl} \not= 0 \wedge \max B_{zl}\not= 0$.
\end{definition}

To better illustrate our definition of closed feature triangles, we select two real-world networks, \textit{Facebook} and \textit{Sinanet}, for empirical analysis. Their node feature types are binary and continuous-valued, respectively. More detailed descriptive information about them will be introduced in the experimental section. 

As shown in Tables~\ref{tab41} and~\ref{tab42}, in term of \textit{Facebook}, the ratio of the number of closed topological triangles within all communities in ground-truths to the total number of closed topological triangles in ground-truths is as high as 93\%, which means that considering closed topological triangles is helpful to improve the performance of community detection. As for closed feature triangles, the number is huge because the probability of formation is much larger than that of closed topological triangles. However, closed feature triangles are not all conducive to community detection. We further examine the number of edges in closed feature triangles within all communities. The results show that most of closed feature triangles contain only zero or one edge, and the number of closed feature triangles containing two or three edges is roughly in the same order of magnitude as the number of closed topological triangles. This inspires us to construct closed feature triangles around adjacent nodes, which also seems to be more in line with the fact that communities are mesoscopic structure~\cite{fortunato2016community}.

In term of \textit{Sinanet}, its feature type is continuous, and we count the number of closed feature triangles based on Definition~\ref{def31}. It is not difficult to find from Tables~\ref{tab41} and~\ref{tab42} that the number of closed topological triangles within all communities is relatively small, while the number of closed feature triangles is still huge. This further reflects the significance of node features as a supplement. Furthermore, the conclusions presented on the number of edges within closed feature triangles are consistent with \textit{Facebook}, which demonstrates that our definition is appropriate. To sum up, we believe that the influence of closed feature triangles formed by adjacent nodes is equal to that of closed topological triangles, and rewrite $WCC(v_i, C_k)$ to $WCC^*(v_i, C_k)$ as follows:
\begin{equation}
\label{equ41}
	\left\{
	\begin{split}
		&\frac{tf({\vv}_{\ii}, {\CC}_{k})}{tf({\vv}_{\ii}, NC)}{\cdot}
		\frac{vtf({\vv}_{\ii}, NC)}{|{\CC}_{k} - \{{\vv}_{\ii}\}| + vtf({\vv}_{\ii}, N(v_i))},
		&tf({\vv}_{\ii}, NC) \neq 0; \\
		&0,&tf({\vv}_{\ii}, NC) = 0; \\
	\end{split}
	\right.
\end{equation}
where $NC = {N(v_i)\cup{\CC}_{k}}$ represents a set consisting of the neighbors of a given node and nodes within the candidate community. $tf(\cdot,\cdot)$ represents the number of closed topological and feature triangles, and $vtf(\cdot,\cdot)$ represents the number of nodes forming at least one closed topological or feature triangle. The rest have been explained before and will not be repeated here.

\subsection{Two-Level Constraints}
Intuitively, if only the network topology is considered, a good community should ensure that internal links are as dense as possible, or form more closed triangles. Extensive research~\cite{DBLP:journals/tkde/RezaeiF16, DBLP:journals/csur/ChakrabortyDMG17} proposes quality metrics such as modularity $Q$~\cite{newman2006modularity} and normalized cut~\cite{DBLP:journals/pami/ShiM00} based on such a principle, and achieves community detection by maximizing or minimizing the value of the given quality metric. However, it is not true that the optimization of quality metrics always performs satisfactorily, e.g., the resolution limit of modularity~\cite{fortunato2007resolution}, the unbalanced scale of communities~\cite{DBLP:journals/tsmc/GaoWZCQ22}, the free-rider effect~\cite{DBLP:journals/pvldb/WuJLZ15}. Although the value of the corresponding quality metric is optimal, the natural community structure of a network is not clearly revealed.

In addition, in network representation, node features are the most important dimension besides the topology structure. Those approaches~\cite{falih2018community, DBLP:journals/csr/Chunaev20} that utilize node features to detect communities usually formulate distance or similarity functions, or consider feature embeddings to evaluate the differences between node features. As a result, nodes in the same community share more features. However, each node has multiple features, not all of which are helpful in determining the community a node belongs to. In some cases, node features even provide contradictory information with topological structure~\cite{DBLP:journals/csr/Chunaev20}. Moreover, existing evaluation methods are generally difficult to apply to different feature types, and too many features also affect the efficiency of evaluation. 

Hence, to identify more natural communities from an attributed network, we cannot blindly pursue the best value for a given quality metric, but at the same time, we should prevent communities from being over-scaled and over-featured. We now describe the proposed two-level constraints from the perspective of tightness and homogeneity, which can alleviate these restrictions mentioned above when $WCC^*$ is used as an optimization objective.

The tightness constraint emphasizes the topology of each community by focusing on the connections between nodes. In general, nodes with higher degrees can be regarded as leaders with greater influence, thus identifying their communities, and the belongings of the remaining nodes will be continuously determined~\cite{fortunato2016community}. However, once the leaders' communities are biased, the impact on the final outcome is severe. We intuitively consider the number of edges between nodes. It is worth noting that the more neighbor nodes of a node belong to the same community, the more likely the node itself belongs to that community. Therefore, we formulate the first-level constraint item as follows:
\begin{definition}
	The tightness constraint of a node $v_i$ and a community $C_k$ is defined as follows:
	\begin{equation}\label{equ:tig}
		T(v_i, C_k) = \frac{\sum_{v_j \in C_k}E_{v_i,v_j}^{in}}{d_i~{\cdot}~|C_k|}.
	\end{equation}
	\label{def32}
\end{definition}

When there is an edge between node $v_i$ and node $v_j \in C_k$, $E_{v_i,v_j}^{in} = 1$, otherwise 0. In other words, the more edges between node $v_i$ and nodes in community $C_k$, the larger the value of $\sum_{v_j \in C_k}E_{v_i,v_j}^{in}$. Thus, by maximizing $T(v_i, C_k)$, the joined node will improve the topological properties of the community. Meanwhile, the degree $d_i$ of node $v_i$ is introduced for normalization, and the scale $|C_k|$ of community $C_k$ is considered to prevent oversize, so that the detected communities are more compact.

The homogeneity constraint grasps the feature distribution of a community by highlighting the similarity between node features. Element-by-element matching~\cite{DBLP:journals/tkde/RezaeiF16} is the simplest method, such as jaccard and cosine similarity, the higher the matching degree, the higher the similarity. Regularization~\cite{DBLP:journals/jnca/JavedYLQB18} is also a widely adopted method, such as L1 and L2 norm, which can mitigate the negative impacts of outliers and missing data. Since we consider both binary and continuous-valued features, the applicability of a given method is our primary concern. Therefore, we formulate the second-level constraint item as follows:
\begin{definition}
	The homogeneity constraint of a node $v_i$ and a community $C_k$ is defined as follows:
	\begin{equation}\label{equ:tig}
		H(v_i, C_k) = \frac{\sum_{v_j \in C_k}{|\mathbf{\ff}_{i} - \mathbf{\ff}_{j}|}}{p~{\cdot}~|C_k|}.
	\end{equation}
	\label{def33}
\end{definition}

When the features of two nodes are highly consistent, the two nodes are strongly similar, and the corresponding value of $|\mathbf{\ff}_{i} - \mathbf{\ff}_{j}|$ will be small regardless of the feature type. Thus, by minimizing $H(v_i, C_k)$, finding nodes that agree on more features, the feature distribution of each community is more explicit. As for feature dimension $p$ and community scale $|C_k|$, their effects are equivalent to $d_i$ and $|C_k|$ in the tightness constraint, so the detected communities are more appropriate.

\noindent\textbf{Discussion:} In general, a well-separated community contains about 100 nodes, and meaningful larger communities can be generated by merging these relatively small communities~\cite{fortunato2016community, DBLP:journals/csr/Chunaev20}. However, unrestricted merging will generate oversized communities, which will further lead to unbalanced communities and free-rider effect. Our quality metric considers higher-order details from the perspective of closed topological and feature triangles, which helps to extract tight core communities, but nodes and edges that are difficult to form closed triangles are excluded from the communities. Our two-level constraints consider lower-order details from the perspective of edge tightness and node feature homogeneity, which guide the growth of core communities as nodes and edges are selectively added. As a consequence, communities remain self-contained unless highly correlated. Therefore, the merging of communities will be regulated and the scale of communities will be widely distributed. 

\subsection{Local Search Framework}
Combining the proposed quality metric with two-level constraints, denoted as $U$, the optimization objective function we maximize in this study is as follows:
\begin{equation}
	\label{equ35}
	U = \sum_{k=1}^{K}\sum_{\forall v_i \in C_k}[WCC^*(v_i, C_k) + T(v_i, C_k) - H(v_i, C_k)].
\end{equation}

When $WCC^*(v_i, C_k) \not= 0$, $U$ is equivalent to the revision of $WCC^*(C)$, which enhances its applicability in different tasks. Otherwise, the optimization objective is reduced to two-level constraints, and the performance of community detection can still be used as a benchmark.

Greedy objective function maximization reduces the computational cost significantly, but the quality of the results highly depends on the order of processing~\cite{DBLP:conf/kdd/ZheSX19}. To prevent this issue, we design a local search framework based on maximizing the utility function of each node extracted from the optimization objective $U$. Specifically, given a node $v_i$, its utility in the community $C_k$ is defined as follows:
\begin{equation}
	\label{equ36}
	u_{ik} = WCC^*(v_i, C_k) + T(v_i, C_k) - H(v_i, C_k).
\end{equation}

Based on Equation~(\ref{equ36}), the utility gain ($\Delta{u}_{ik \rightarrow ik'}$) of node $v_i$ moving from the current $C_k$ to a new community $C_{k'}$ can be measured. Thus, our search process for both non-overlapping and overlapping communities is summarized as follows:

\begin{itemize}
	\item{\textbf{Step~1: (Cumulative Node Utility Updating).} 
		Let's denote $\overline u_{ik}^{t}$ as the cumulative utility of node $v_i$ in round $t$, which also reflects the degree to which node $v_i$ belongs to community $C_k$ at this time. Given any community $C_{k'}$ that includes at least one neighbor of $v_i$, then in round $t+1$, the cumulative utility of $v_i$ joining the community ($\overline u_{ik'}^{t+1}$) is updated as follows: 
		\begin{equation}\label{equ38}
			\overline u_{ik'}^{t+1} = \alpha~{\cdot}~\Delta{u}_{ik \rightarrow ik'} + (1-\alpha)~{\cdot}~\overline u_{ik}^t.
		\end{equation}
		Note that the value of $\alpha \in [0, 1]$ indicates a trade-off between historical utility and current utility gain. We empirically specify it as 0.2.}
	\item{\textbf{Step~2: (Candidate Community Labels Filtering).}
		Node $v_i$ filters its candidate community labels set ($CL_{i}^{t+1}$) for round $t+1$ based on the cumulative utility collected from Step~1:
		\begin{equation}\label{equ39}
			CL_{\ii}^{\ttt+1} = \{\kk'~|~\overline u_{ik'}^{t+1} \geq \overline u_{ik}^t\}.
		\end{equation}
		Note that the community for $\vv_\ii$ in round $\ttt+1$ is either its current community $C_k$ or the community $C_{k'}$ in $CL_{\ii}^{\ttt+1}$.}
	
	\item{\textbf{Step~3: (Appropriate Community Labels Assigning).}
		1) Sort the community labels in $CL_{\ii}^{\ttt+1}$ based on successive decreasing of $\overline u_{ik'}^{t+1}$, and remove the last $1/|CL_{\ii}^{\ttt+1}|$~\cite{DBLP:journals/tkde/LuZQK19} community labels; 
		2) The remaining community labels in $CL_{\ii}^{\ttt+1}$ are regarded as overlapping communities to which $v_i$ belongs in round $t+1$;
		3) Among them, the community label with the greatest utility is regarded as the non-overlapping community to which $v_i$ belongs in round $t+1$.}
\end{itemize}

Algorithm~\ref{alg1} shows the detailed procedures of the above process step by step. Our framework only takes $G$ as input. The initial configuration (line 1) is to assign each node a unique community label. After that, the program loops in four steps. Specifically, in lines 3–7, each node updates the cumulative utility, filters candidate community labels, and determines new communities based on different types of community detection; in lines 8–10, each node selects the community label with the greatest cumulative utility for the next round of search. Finally, if no nodes have changed community affiliation or satisfied the convergence criterion ($t > 20$), the above procedures will terminate and output non-overlapping and overlapping partitions.

\begin{algorithm}[t!]
	\caption{Local Search Framework (LSF)}
	\label{alg1}
	\begin{algorithmic}[1]
		\REQUIRE{An attributed network $\GG =(\VV, \EE, \FF)$;}
		\ENSURE{Non-overlapping (overlapping) partition $\CC$;}
		
		\STATE $\ttt=0$, initialize each node belongs to a community:\\
		$\forall \vv_{\ii}\in{\VV}$, $k{\gets}\ii$, $CL_{\ii}^{0}{\gets}\{k\}$, ${\overline u_{ik}^{0}}{\gets}0$\\
		$k'{\gets}$any community $C_{k'}$ where ${v_i}'$s neighbors belong\;
		
		\WHILE{$C$ changed in the previous round or $\ttt \leq 20$}
		\FOR{each node $\vv_{\ii} \in \VV$}
		\STATE Cumulative Node Utility Updating\;
		\STATE Candidate Community Labels Filtering\;
		\STATE Appropriate Community Labels Assigning\;
		\ENDFOR
		\FOR{each node $\vv_{\ii} \in \VV$}
		\STATE $k{\gets}$community with the greatest cumulative utility\;
		\ENDFOR
		
		\STATE $\ttt{\gets}\ttt+1$\;
		\ENDWHILE
		\STATE return $C$\;
	\end{algorithmic}
\end{algorithm}

\section{Experiments}\label{sect5}
In this section, we evaluate the overall performance of our proposed framework, LSF, on seven real-world network datasets. All experiments were performed on a PC equipped with an Intel quad-core i7 processor (2.60 GHz) and 16GB memory. We conduct experiments as follows: 

(1) By comparing the overall performance of LSF and twelve baseline approaches, we demonstrate the effectiveness of the proposed framework. (Section~\ref{sect52})

(2) By investigating the running time of LSF and twelve baseline approaches, we verify the scalability and efficiency of the proposed framework. (Section~\ref{sect53})

(3) By revealing the convergence of LSF, the effect of two-level constraints, and the topological density and feature entropy of each found community, we make an in-depth analysis of the proposed framework. (Section~\ref{sect54})

\subsection{Experimental setup}\label{sect51}
\noindent\textbf{Experimental data.} Seven real-world network datasets are used in our experiments, and their ground-truth communities are all known. \emph{Facebook}, \emph{Twitter}, and \emph{Gplus} are friendship networks derived from online social networking sites, and the profile of each node is described by binary features. \emph{Youtube} and \emph{Livejournal} are friendship networks obtained from a video sharing site and a free blogging site, and nodes have no feature details. These five networks are available from the Stanford Network Analysis Platform (SNAP{\footnote{http://snap.stanford.edu/data/index.html.}}). \emph{Sinanet} is a microblog user relationship network extracted from the sina-microblog site. The topic distribution of each user in the forums generated by the LDA topic model is treated as a continuous-valued feature for each node. \emph{Diabetes} is a citation network formed by citation relationships for scientific publications on diabetes from the PubMed database. The continuous-valued feature for each node corresponds to the TF/IDF weighted word representation of each publication. These two networks are provided by GitHub{\footnote{https://github.com/smileyan448/Sinanet.}} and LINQS{\footnote{https://linqs.soe.ucsc.edu/data.}}, respectively.

\begin{table*}[t!]
	\centering
	\caption{Network Datasets used in Experiments}
	\begin{tabular}{ l | c c c c c c}
		\hline
		Network	&$|\VV|$		&$|\EE|$	&$|\FF|$	&$\left\langle{\dd}\right\rangle$	&$\#Comm.$	&$Overlaps$\\
		\hline
		\emph{Sinanet}	&3, 490	&28, 657	&10	&16.42	&10	&1.00\\
		\emph{Diabetes}	&19, 717	&44, 338	&500	&4.50	&3	&1.00\\
		\hline
		\emph{Facebook} &4, 039	&88, 234	&157	&43.69	&146	&1.46\\
		\emph{Twitter}	&81, 306	&1, 342, 296	&33, 208	&33.02	&3, 170	&2.22\\
		\emph{Gplus}	&102, 100	&12, 113, 501	&805	&237.30	&438	&2.69\\
		\hline
		\emph{Youtube}	&1, 134, 890	&2, 987, 624	&-	&5.27	&8, 385	&2.40\\
		\emph{Livejournal}	&3, 997, 962	&34, 681, 189	&-	&17.35	&287, 512	&5.88\\
		\hline
	\end{tabular}
	\label{tab51}
\end{table*}

We provide some basic statistics for these datasets in Table~\ref{tab51}, where $|\VV|$, $|\EE|$, $|\FF|$, $\left \langle {\dd} \right \rangle =  2|\EE|/|\VV|$, $\#Comm.$, and $Overlaps = \sum^{K^*}_{k^*=1}|C^*_{k^*}|/|V|$ indicate the number of nodes, edges, and features, the average degree of the network, the number of ground-truth communities, and the overlap rate of the ground-truth partition, respectively, where $K^*=\#Comm.$ and $C^*_{k^*}$ is the $k^*$th community in the ground-truth partition.

\noindent\textbf{Baseline approaches.} Twelve state-of-the-art community detection approaches are selected as baselines for performance comparison, and we divide them into three groups. The first group of baselines consider topology and binary features: CESNA~\cite{DBLP:conf/icdm/YangML13} finds overlapping communities based on a probabilistic generative model combining network topology and node features. EDCAR~\cite{DBLP:conf/pakdd/GunnemannBFS13} uses the greedy stochastic adaptive search principle to approximate the optimal clustering solution to detect overlapping clusters with high topology density and high feature similarity. CAMAS~\cite{DBLP:journals/inffus/BuGLC17} is based on the established cluster-aware multi-agent system to achieve overlapping clustering in attributed networks.

The second group of baselines consider topology and continuous-valued features: I-Louvain~\cite{DBLP:conf/ida/CombeLGE15} is implemented following the exploration principle of Louvain and optimizes the modularity and the proposed inertia based modularity. NAGC~\cite{DBLP:journals/corr/abs-1810-00946} performs non-linear attributed network clustering via symmetric non-negative matrix factorization with positive unlabeled learning. $k$NN-enhance~\cite{jia2017node} adds the $k$ nearest neighbor graph of node features to alleviate the sparsity and noise effects of the original network, thus strengthening the found community structure.

As for the remaining baselines, they only focus on topology: SCD~\cite{Prat2014WWW} divides the network into non-overlapping groups by maximizing the weighted community clustering metric. InfoMap~\cite{rosvall2007maps} reveals that non-overlapping communities aim to optimize a quality metric expressing the code length of an infinitely long random walk taking place on the network. SCoDA~\cite{DBLP:journals/corr/HollocouMBL17} randomly and uniformly picks an edge in the network that is more likely to connect two nodes in the same community than two nodes in different communities, and exploits this idea to build non-overlapping communities by local changes at each edge arrival. FOCS~\cite{DBLP:journals/tkde/BandyopadhyayCS15} explores locally well-connected overlapping communities by computing community connectedness and neighborhood connectedness scores for each node. BigClam~\cite{DBLP:conf/wsdm/YangL13} proposes a conceptual model of network community structure, cluster affiliation model, and then employs non-negative matrix factorization to find overlapping communities based on this. LPANNI~\cite{DBLP:journals/tkde/LuZQK19} detects overlapping communities by adopting fixed label propagation sequence based on the ascending order of node importance and label update strategy based on neighbor node influence and historical label preferred strategy.

\noindent\textbf{Parameter settings.} All baseline packages are implemented in C++, Java, or Python and can be found on the websites provided by the papers. Since CESNA, NAGC, $k$NN-enhance, and BigClam require a user-given parameter as the number of communities they found, we set this parameter to $\#Comm.$ for comparison. All other parameters of the selected baselines use their default settings.

\noindent\textbf{Evaluation measures.} We do not make any special distinction between non-overlapping and overlapping community detection. Four evaluation measures are chosen to assess the quality of detected communities: Average F1 Score ($AvgF1$)~\cite{DBLP:conf/wsdm/YangL13}, Modularity $Q$~\cite{newman2006modularity}, Density~\cite{DBLP:journals/csur/ChakrabortyDMG17}, and Entropy~\cite{jia2017node}. Given two partitions $C =\{{C_{1},...,C_{K}}\}$ and $C^* =\{{C^*_{1},...,C^*_{K^*}}\}$, $AvgF1$ is defined as follows:
\begin{align*}
	Avg&F1 = \frac{1}{2K}\sum_{C_{k}\in{C}}\max_{C^*_{k^*}\in{C^*}}F1(C_{k}, C^*_{k^*}) 
	+ \frac{1}{2K^*}\sum_{C^*_{k^*}\in{C^*}}\max_{C_{k}\in{C}}F1(C^*_{k^*},C_{k}),\\ 
	&F1(C_{k},C^*_{k^*}) = 2{\cdot}\frac{Precision(C_{k},C^*_{k^*}){\cdot}Recall(C_{k},C^*_{k^*})}{Precision(C_{k},C^*_{k^*})+Recall(C_{k},C^*_{k^*})},\\
	P&recision(C_{k},C^*_{k^*}) = \frac{|C_{k}{\cap}C^*_{k^*}|}{|C_{k}|},~~Recall(C_{k},C^*_{k^*}) = \frac{|C_{k}{\cap}C^*_{k^*}|}{|C^*_{k^*}|}.\\
\end{align*}

Given a network with $n$ nodes and $m$ edges, and one partition $C =\{{C_{1},...,C_{K}}\}$, modularity and density are defined as follows:
\begin{align*}
	Modularity~Q &= \frac{1}{2m}\sum_{i=1}^n\sum_{j=1,j\not=i}^n(A_{ij} - \frac{d_{i}d_{j}}{2m})\sum_{k=1}^Kc_{ik}c_{jk},\\
	Density~(C_k) &= \frac{2E^{in}_k}{|C_k|\cdot(|C_k|-1)},
\end{align*}
where $c_{ik}$~($c_{jk}$) represents the degree of belonging of node $v_i$~($v_j$) to the $k$th community, and $E^{in}_k$ represents the number of internal edges in the $k$th community.

Given a network with $n$ nodes and $p$ features per node, and one partition $C =\{{C_{1},...,C_{K}}\}$, entropy is defined as follows:
\begin{align*}
	Entropy~(C_k) &= - \frac{|C_k|}{n}\sum_{l=1}^pP_{lk}\log(P_{lk}),
\end{align*}
where $P_{lk}$ is the fraction of nodes with $l$th feature in the $k$th community.

It is expected that better communities (high tightness and homogeneity) of a given network data will have larger values of $AvgF1$, modularity $Q$, density, and smaller value of entropy.

\begin{table*}[t!]
	\centering
	\caption{Effectiveness Evaluation of LSF and Baselines ignoring Node Features ($AvgF1$)}
	\begin{tabular}{l | c c c c c c | c}
		\hline
		Network	&SCD	&InfoMap	&SCoDA	&FOCS	&BigClam	&LPANNI	&LSF\\
		\hline
		\emph{Sinanet}	&0.096	&0.167	&0.025	&0.114	&0.286	&0.174	&\textbf{0.308}\\
		\emph{Diabetes}	&0.004	&0.325	&0.003	&0.098	&0.002	&0.097	&\textbf{0.397}\\
		\hline
		\emph{Facebook} &0.197	&0.405	&0.223	&0.328	&0.422	&0.368	&\textbf{0.452}\\
		\emph{Twitter}	&0.167	&0.151	&0.178	&0.078	&0.101	&0.133	&\textbf{0.299}\\
		\emph{Gplus}	&0.039	&0.106	&0.049	&0.162	&0.093	&NaN	&\textbf{0.261}\\
		\hline
		\emph{Youtube}	&0.177	&0.084	&0.161	&0.148	&0.038	&0.141	&\textbf{0.194}\\
		\emph{Livejournal}	&0.139	&0.059	&\textbf{0.169}	&0.145	&0.103	&NaN	&0.165\\
		\hline
	\end{tabular}
	\label{tab53}
\end{table*}

\begin{table*}[!t]
	\centering
	\caption{Efficiency Evaluation of LSF and Baselines ignoring Node Features (in seconds)}
	\begin{tabular}{l |cccccc| c}
		\hline
		Network	&SCD	&InfoMap	&SCoDA	&FOCS	&BigClam	&LPANNI	&LSF\\
		\hline
		\emph{Sinanet}	&\textbf{0.18}	&0.85	&0.40	&1.02	&4.25	&79.56	&0.65\\
		\emph{Diabetes}	&0.33	&1.06	&0.78	&0.99	&\textbf{0.26}	&14.99	&1.09\\
		\hline
		\emph{Facebook}	&0.53	&1.17	&0.96	&1.21	&19.50	&510.40	&1.16\\
		\emph{Twitter}	&4.69	&19.51	&\textbf{1.44}	&4.26	&467.36	&24, 930.29	&50.24\\
		\emph{Gplus}	&67.53	&95.86	&\textbf{12.35}	&959.38	&1, 804.60	&NaN	&494.67\\
		\hline
		\emph{Youtube}	&17.49	&146.45	&\textbf{5.57}	&39.70	&22, 680.47	&80, 655.41	&100.73\\
		\emph{Livejournal}	&194.18	&1, 553.29	&\textbf{30.40}	&232.65	&37, 426.82	&NaN	&2, 057.95\\
	\end{tabular}
	\label{tab54}
\end{table*}

\subsection{Effectiveness Evaluation}\label{sect52}
\begin{figure}[t!]
	\centering
	\includegraphics[width=0.48\textwidth]{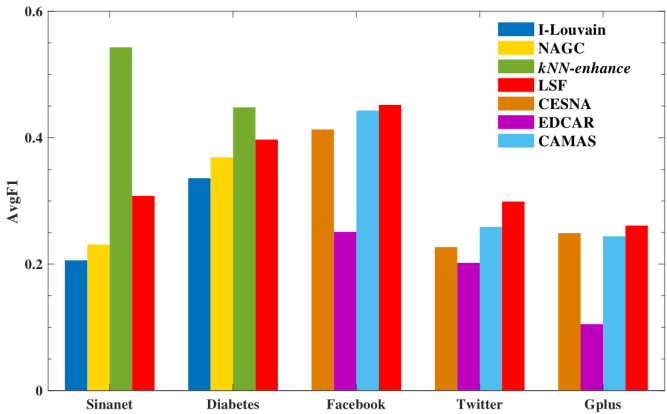}
	\caption{Effectiveness Evaluation of LSF and Baselines using Node Features}
	\label{fig521}
\end{figure}

To better evaluate the effectiveness of our framework, we comprehensively compare the quality of non-overlapping and overlapping communities found by LSF and twelve baselines on seven experimental datasets. Figure~\ref{fig521} and Table~\ref{tab53} present the $AvgF1$ scores for all tested approaches, with the best scenarios in bold. The NaNs in Table~\ref{tab53} are due to the fact that the node's feature type is not suitable for the baseline tool, or some baseline tools run out of memory or too expensive to run as the scale of the network increases.

We have the following observations and conclusions from these results. In terms of approaches that consider continuous-valued features, $k$NN-enhance performs the best, which indicates that our processing of continuous-valued features still needs to be further improved. However, the performance of our framework is also acceptable as it outperforms I-Louvain and NAGC. In terms of approaches that consider binary features, although CESNA and CAMAS are competitive, our framework performs significantly better than all of them, which also demonstrates that our processing of binary features is appropriate. In terms of approaches that only focus on topology, our framework maintains the best performance, and only SCoDA is slightly better than ours on \emph{Livejournal}. This declares that the integration of network topology and node features is indeed helpful to improve the quality of the found communities. 

It is not difficult to find that the superiority of our framework is obvious. Specifically, it can handle different types of node features and exhibits satisfactory overall performance in different community detection tasks.

\begin{figure}[t!]
	\centering
	\includegraphics[width=0.48\textwidth]{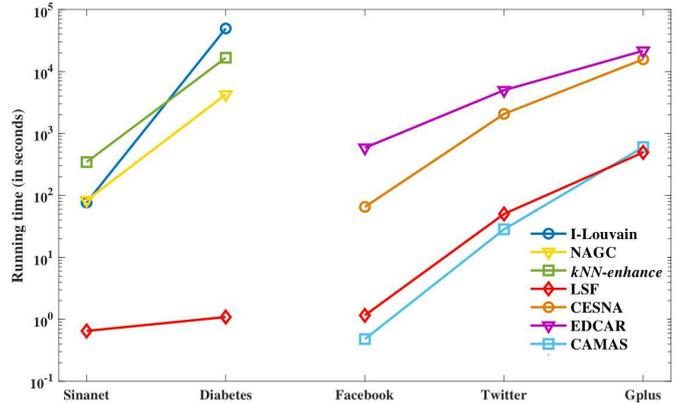}
	\caption{Efficiency Evaluation of LSF and Baselines using Node Features}
	\label{fig531}
\end{figure}

\subsection{Efficiency Evaluation}\label{sect53}
To better evaluate the efficiency and scalability of our framework, we present the running time of LSF and twelve baselines on seven experimental datasets in Figure~\ref{fig531} and Table~\ref{tab54}, with the best scenarios also in bold. Like Table~\ref{tab53}, the NaNs in Table~\ref{tab54} are also due to the fact that the node's feature type is not suitable for the baseline tool, or some baseline tools run out of memory or too expensive to run as the scale of the network increases.

We have the following findings and determinations from these results. In terms of approaches that consider continuous-valued features, the superiority of our framework is obvious. Although the performance of $k$NN-enhance is better than ours, its time cost is too high to be applied to larger-scale networks. In terms of approaches that consider binary features, CAMAS is the most efficient, followed by our framework, and both consistently outperform CESNA and EDCAR. The efficiency difference between ours and CAMAS tends to weaken as the scale of the network increases. In terms of approaches that only focus on topology, SCoDA is the best, followed by SCD and FOCS, then InfoMap and ours, and finally BigClam and LPANNI. Since our framework considers both topology and node features, our efficiency is inferior to some baselines, but the overall performance is still acceptable.

It is not difficult to find that approaches that combine topology and node features are generally less efficient than approaches that only focus on topology. However, our framework maintains satisfactory efficiency and scalability, on the same order of magnitude as CAMAS and InfoMap. While still not as good as some baselines, we can identify higher quality communities, which is enough to make up the slight inferiority on efficiency.

\subsection{Inside Analysis}\label{sect54}
To better interpret the good performance of our framework, we here take a further look inside the framework. Specifically, we analyze the convergence of LSF, the effect of the proposed constraints, and the quality of each community found.

\begin{figure}[t!]
	\centering
	\includegraphics[width=0.48\textwidth]{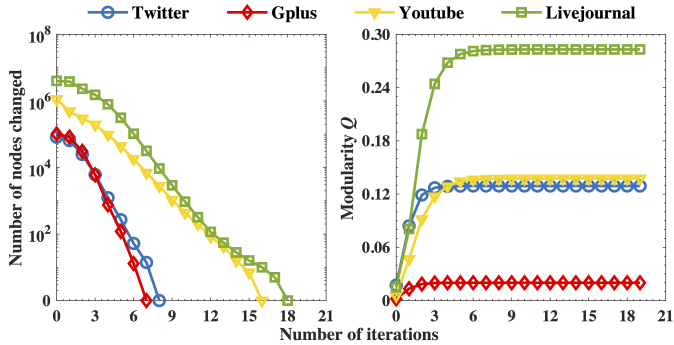}
	\caption{Convergence Verification of LSF based on the Number of Nodes Changed and Modularity $Q$}
	\label{fig541}
\end{figure}

\noindent\textbf{Verification of convergence.} We select four datasets \emph{Twitter}, \emph{Gplus}, \emph{Youtube}, and \emph{Livejournal} as representatives, and examine the number of nodes whose community labels change and the value of modularity $Q$ in each round of our framework runs. The results are shown in Figure~\ref{fig541}.

We can find: In terms of the number of nodes whose community labels change, \emph{Twitter} and \emph{Gplus} have similar trends, with the number of nodes decreasing by an order of magnitude in almost every round, while \emph{Youtube} and \emph{Livejournal} have similar trends, generally two or three rounds will also decrease by an order of magnitude. In terms of the value of modularity $Q$, the trends are very similar regardless of the datasets. The $Q$ value increases continuously over several rounds and then tends to be smooth. In summary, our framework is robust and converges fast, and exhibits great potential on large-scale networks. Considering the fact that \emph{Twitter} and \emph{Gplus} are dense, and \emph{Youtube} and \emph{Livejournal} are sparse, the results of the convergence verification reveal that our framework seems to be more suitable for dense networks.

\begin{figure}[t!]
	\centering
	\includegraphics[width=0.47\textwidth]{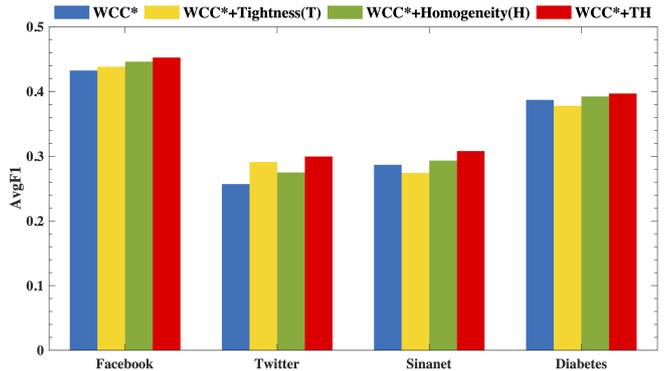}
	\caption{Evaluation of the Effects of Two-level Constraints}
	\label{fig542}
\end{figure}

\noindent\textbf{Effects of two-level constraints.} We select four datasets \emph{Facebook}, \emph{Twitter}, \emph{Sinanet}, and \emph{Diabetes} as representatives, and take an ablation study to investigate the effects of the proposed constraints. Specifically, we consider four cases: without the tightness and homogeneity constraints, with the tightness or homogeneity constraint, and with the tightness and homogeneity constraints. The results are shown in Figure~\ref{fig542}.

We can find: When the network is dense (\emph{Facebook} and \emph{Twitter}), the effect of the tightness constraint is positive and the performance of our framework improves. The effect of the homogeneity constraint is unstable, and a few features further enhance the degree of improvement, but as the number of features increases, the effect of the homogeneity constraint decreases or even becomes negative. When the network is sparse (\emph{Sinanet} and \emph{Diabetes}), the effect of the tightness constraint is negative and the performance of our framework generally declines. The effect of the homogeneity constraint is positive, but also seems to be controlled as the number of features increases. In summary, it is indeed useful with two-level constraints than without. However, if only one constraint is considered, due to the heterogeneity of network topology and node features, the overall performance improvement depends on the trade-off between the density of the topology and the number of node features.

\begin{figure}[t!]
	\centering
	\includegraphics[width=0.48\textwidth]{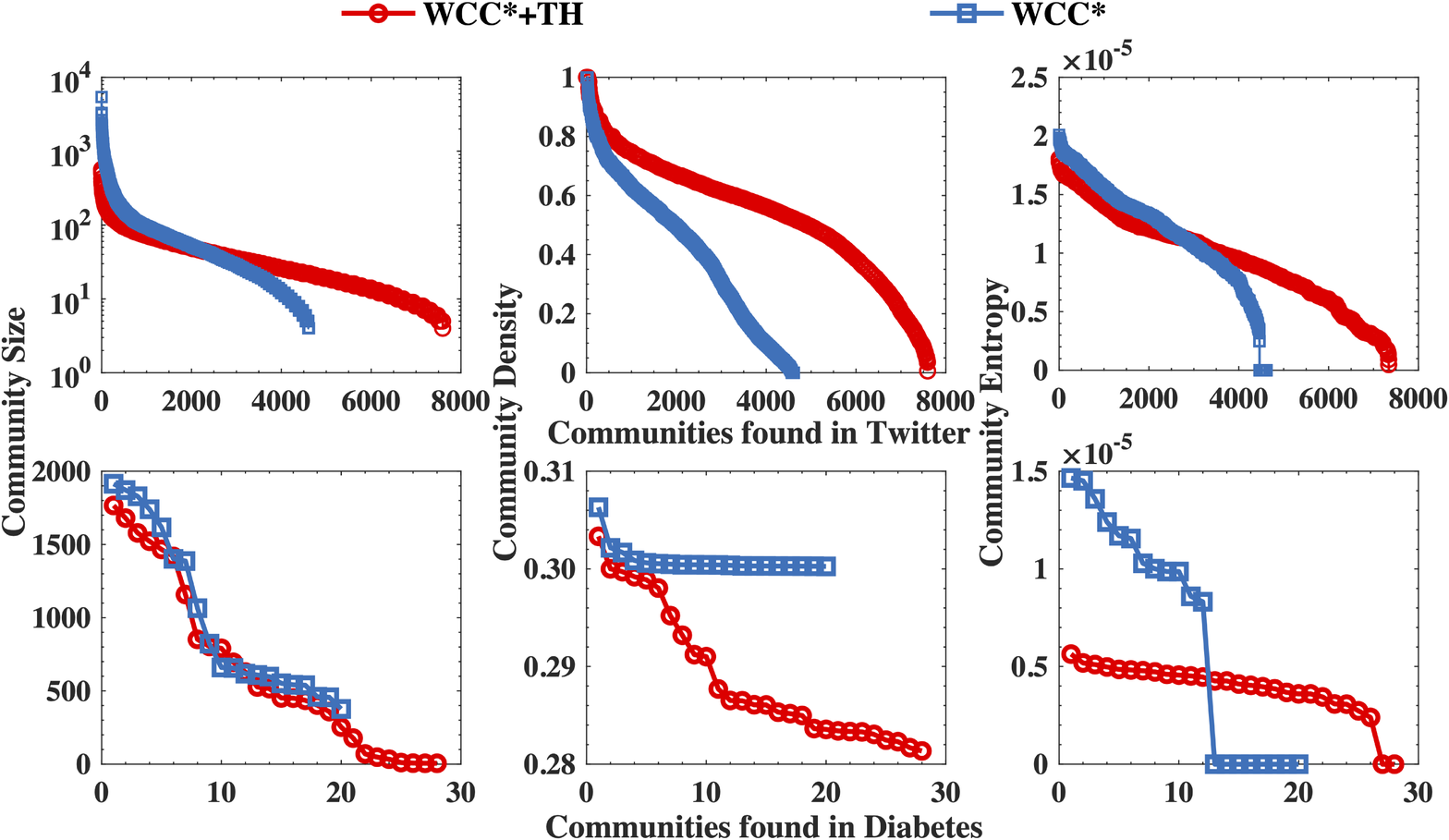}
	\caption{Evaluation of Topological Density and Feature Entropy for Each Community Found}
	\vspace{-0.1cm}
	\label{fig543}
\end{figure}

\noindent\textbf{Analysis of found communities.} We select two datasets \emph{Twitter} and \emph{Diabetes} as representatives, and further explore the topological density and feature entropy scores for each community found with and without two-level constraints. The results are shown in Figure~\ref{fig543}.

We can find: In terms of tightness, larger communities are more likely to contain more edges and have a larger density score. \emph{Twitter} is dense, and the tightness constraint makes it easier for closely connected nodes to be in the same community, thus improving the quality of each community. In contrast, \emph{Diabetes} is sparse, and the tightness constraint forces weakly connected nodes to the same community, and the overall performance is compromised. In terms of homogeneity, smaller communities tend to exhibit more uniform node features and have a smaller entropy score. \emph{Twitter} has too many features, and the homogeneity constraint considers the differences in each dimension of node features, thus reducing the quality as the community scale becomes smaller. On the contrary, \emph{Diabetes} has a small number of features and is the relatively smooth continuous-valued type. The homogeneity constraint guarantees the feature distribution of each community, and the overall performance is improved. In summary, by balancing the proportions of topology tightness and feature homogeneity, our searched communities are more compact and reasonable, resulting in better performance.

\section{Conclusions and Future Work}\label{sect6}
In this study, we first introduce a novel quality metric based on the concepts of closed topology and feature triangles, which not only evaluates the quality of the found communities, but can also be treated as an optimization objective function. We then formulate two constraint items from the perspective of node features and network topology, which alleviate unbalanced communities and free-rider effect by regulating the topological tightness and feature homogeneity of each community found. Finally, combining the proposed metric and constraint items as the objective function, we develop a local search framework by optimizing it to achieve community detection in attributed networks. Experimental results on seven real-world networks show that the overall performance of our framework consistently outperforms all selected state-of-the-art approaches. In the future, we will focus on the dynamic evolution of the community structure and investigate how to design parallel schemes to make our framework more efficient.

\section*{Acknowledgements}
This work was supported in part by the Key Program of National Natural Science Foundation of China under grant 92046026, and in part by Jiangsu Provincial Policy Guidance Program under grant BZ2020008, and in part by the Key Discipline Construction Project of Cyberspace Security of the 14th Five-Year Plan of Jiangsu Province, and in part by the High-Level Introduction of Talent Scientific Research Start-up Fund of Jiangsu Police Institute under Grant JSPI21GKZL401, and in part by the Philosophy and Social Foundation of the Jiangsu Higher Education Institutions of China under Grant 2019SJA0443.




\bibliographystyle{elsarticle-num}
\biboptions{sort&compress}
\bibliography{myRefs}





\end{document}